\documentclass[a4paper]{article}

\usepackage{INTERSPEECH2022}
\usepackage{cite}

\title{Tree-constrained Pointer Generator with Graph Neural Network Encodings for Contextual Speech Recognition}
\name{Guangzhi Sun, Chao Zhang, Philip C. Woodland\thanks{Guangzhi Sun is funded by Cambridge Trust.}}
\address{
  Cambridge University Engineering Dept., Trumpington St., Cambridge, CB2 1PZ U.K.}
\email{\texttt{\{gs534,cz277,pcw\}@eng.cam.ac.uk}}

\begin{document}

\maketitle
\begin{abstract}
Incorporating biasing words obtained as contextual knowledge is critical for many automatic speech recognition (ASR) applications. This paper proposes the use of graph neural network (GNN) encodings in a tree-constrained pointer generator (TCPGen) component for end-to-end contextual ASR. By encoding the biasing words in the prefix-tree with a tree-based GNN, lookahead for future wordpieces in end-to-end ASR decoding is achieved at each tree node by incorporating information about all wordpieces on the tree branches rooted from it, which allows a more accurate prediction of the generation probability of the biasing words. Systems were evaluated on the Librispeech corpus using simulated biasing tasks, and on the AMI corpus by proposing a novel visual-grounded contextual ASR pipeline that extracts biasing words from slides alongside each meeting. Results showed that TCPGen with GNN encodings achieved about a further 15\% relative WER reduction on the biasing words compared to the original TCPGen, with a negligible increase in the computation cost for decoding.
\end{abstract}

\section{Introduction}

Contextual speech recognition aims at addressing the long-tail word problem in end-to-end ASR systems by incorporating contextual knowledge, and has become increasingly important in many applications \cite{shallow_context_1,shallow_context_2,shallow_context_3,deep_context_1,deep_context_2,deep_context_3,deep_context_4,deep_context_5,deepshallow,DBRNNT,dialogue_contextual,dialogue_contextual_2,audiovisual_context,lm_pointer,unsupervised_context,word_mapping}. Informative contextual knowledge can be extracted from text-based resources such as a users' contact book, playlist and the ontology in a dialogue system, or from other modalities such as presentation slides or objects in a scene. Such contextual knowledge is often integrated via contextual biasing which represents contextual knowledge as a \textit{biasing list} of words (referred to as \textit{biasing words}) that are likely to appear in a given context. Despite the relatively small impact of biasing words on the overall word error rate (WER), biasing words are often highly valuable content words, as they are nouns or proper nouns and are indispensable to downstream understanding tasks. Meanwhile, these words are more likely to be correctly recognised if they are included in the biasing list. Therefore, contextual biasing is crucial to the correct recognition of these words in an end-to-end ASR system.

Incorporating dynamic contextual knowledge into end-to-end ASR systems is a challenging problem. Dedicated contextual biasing approaches have been developed, such as shallow fusion (SF) with a special weighted finite-state transducer (WFST), a language model (LM) adapted for contextual knowledge \cite{shallow_context_1,shallow_context_2,shallow_context_3,lm_pointer,unsupervised_context,word_mapping}; attention-based deep context approaches \cite{deep_context_1,deep_context_2,deep_context_3,deep_context_4,deep_context_5}, and also deep biasing (DB) with a prefix-tree for improved efficiency when dealing with large biasing lists \cite{deepshallow,DBRNNT}. More recently, model components that use neural shortcuts to directly modify ASR output distributions were proposed for contextual biasing \cite{TCPGen, MEM, neuralfst}. 

Further exploiting the prefix-tree structure in the tree-constrained pointer generator component (TCPGen) \cite{TCPGen} for contextual biasing, this paper proposes the use of graph neural network (GNN) encodings in TCPGen to obtain more powerful node representations for the prefix-tree. Specifically, a tree recursive neural network (tree-RNN) \cite{RNN1, RNN2}, which is an instance of a GNN, is used to obtain a lookahead functionality for prefix-tree search in end-to-end ASR decoding. Each node representation, encoded with a tree-RNN, encapsulates information about not only its corresponding wordpiece but also all wordpieces on its branches. 
The use of such node representations in TCPGen improves the prediction of the generation probability for the biasing words to better determine how much contextual biasing is needed by incorporating information about future wordpieces for lookahead at each ASR decoding step.

TCPGen with GNN encodings is integrated into both the attention-based encoder-decoder (AED) model and the recurrent neural network transducer (RNN-T) model in this paper. Experiments were first performed on the Librispeech audiobook data using simulated contextual knowledge, where biasing lists were organised for each utterance following \cite{DBRNNT, TCPGen}, as it was shown to be a valid simulation. To further demonstrate the effectiveness of TCPGen and GNN encodings in real-life scenarios, a visual-grounded contextual ASR pipeline is also proposed for the augmented multi-party interaction (AMI) meeting data, which obtains the possible biasing words from the presentation slides alongside each meeting with an optical character recognition (OCR) tool. 
Consistent improvements in WER were achieved in both AED and RNN-T using GNN encodings in TCPGen compared to the original TCPGen system, with a significant WER reduction on the biasing words in particular. 

The rest of this paper is organised as follows: Sec.~\ref{sec:relwork} reviews related studies. Sec.~\ref{sec:TCPGen} introduces TCPGen, followed by Sec.~\ref{sec:RecTCPGen} which gives details about GNN encodings. Sec.~\ref{sec:setup} describes the experimental setup, and Sec,~\ref{sec:result} discusses the results. Finally, conclusions are provided in Sec.~\ref{sec:conclusion}.

\section{Related work}
\label{sec:relwork}
\subsection{End-to-end contextual speech recognition}
\vspace{-0.1cm}
Previous research on end-to-end contextual speech recognition has explored both SF-based score-level interpolation and deep neural representation approaches. For SF-based methods \cite{shallow_context_1,shallow_context_2,shallow_context_3,word_mapping}, biasing lists are represented as extra WFSTs incorporated into a class-based LM, which usually relies on special context prefixes like ``call'' or ``play''. Meanwhile, neural network-based deep context approaches \cite{deep_context_1,deep_context_2,deep_context_3,deep_context_4,deep_context_5} address the dependency issue on fixed syntactic prefixes, but becomes more memory intensive and less effective for large biasing lists. Work in \cite{deepshallow, DBRNNT} jointly used deep context and SF in an RNN-T, along with a prefix-tree search for improved efficiency. It also proposed a simulation of biasing tasks using a public dataset that was adopted in \cite{TCPGen,contextselection}. More recently, neural shortcuts to the final output distribution via a pointer generator \cite{TCPGen, MEM} or neural-FST \cite{neuralfst} have been proposed which can be optimised in an end-to-end fashion. The TCPGen \cite{TCPGen} approach further improved the efficiency by combining the neural shortcut with the symbolic prefix-tree search to handle large biasing lists.

\vspace{-0.1cm}
\subsection{Tree-RNN}
\vspace{-0.1cm}
The tree-RNN \cite{RNN1,RNN2} has been widely applied to various types of tree-structured data. In language technologies, sentences can be arranged into tree structures according to their syntax, to be encoded by the tree-RNN or its variants. Such an encoding has been applied for semantic or sentiment classification \cite{RNNsemantic1, RNNsemantic2, RNNsentiment1}, named-entity recognition \cite{RNNNER}, and has also been used in neural machine translation \cite{RNNtranslation1, RNNtranslation2}. Moreover, encoding of subword-unit-based tree structures using tree-RNN \cite{RNNsubword,RNNsubword2} has also been studied for better word representations.

\section{Tree-constrained pointer generator}
\label{sec:TCPGen}
TCPGen is a neural network-based component combining the symbolic prefix-tree search with a neural pointer generator \cite{pointer_1} for contextual biasing, which enables end-to-end optimisation with ASR systems including AED \cite{e2e_aed_1} and RNN-T \cite{e2e_rnnt_1}. At each output step, TCPGen calculates a distribution over all valid wordpieces constrained by a word-piece-level prefix tree built from the biasing list (referred to as the TCPGen distribution). TCPGen also predicts a generation probability indicating how much contextual biasing is needed at a specific output step. The final output distribution is thus the interpolation between the TCPGen distribution and the original AED or RNN-T output distribution, weighted by the generation probability.

Specifically, a set of valid wordpieces, $\mathcal{Y}^\text{tree}_i$, is obtained by searching the prefix-tree with a given history output sequence. Then, denoting $\mathbf{x}_{1:T}$ and $y_i$ as input acoustic features and output wordpieces, $\mathbf{q}_i$ as the query vector carrying the history and acoustic information, and $\mathbf{K}=[...,\mathbf{k}_j,...]$ as the key vectors, a scaled dot-product attention is performed between $\mathbf{q}_i$ and $\mathbf{K}$ to compute the TCPGen distribution $P^\text{ptr}$ and an output vector $\mathbf{h}^{\text{ptr}}_i$ as shown in Eqn. \eqref{eq:TCPGen_attention} and Eqn. \eqref{eq:TCPGen_value}.
\vspace{-0.1cm}
\begin{equation}
    P^{\text{ptr}}(y_{i}|y_{1:i-1},\mathbf{x}_{1:T}) = \text{Softmax}(\text{Mask}(\mathbf{q}_i\mathbf{K}^\text{T}/\sqrt{d})),
    \label{eq:TCPGen_attention}
    \vspace{-0.1cm}
\end{equation}
\begin{equation}
    \mathbf{h}^{\text{ptr}}_i = \sum\nolimits_{j} P^{\text{ptr}}(y_i=j|y_{1:i-1},\mathbf{x}_{1:T})\,\mathbf{v}^\text{T}_j,
    \label{eq:TCPGen_value}
    \vspace{-0.1cm}
\end{equation}
where $d$ is the size of $\mathbf{q}_i$ (see \cite{transformer}), Mask$(\cdot)$ sets the probabilities of wordpieces that are not in $\mathcal{Y}^{\text{tree}}_i$ to zero, and $\mathbf{v}_j$ is the value vector relevant to $j$. In AED, the query combines the context vector and the previously decoded token embedding, while the keys and values are computed from the decoder wordpiece embedding, with a shared projection matrix. The generation probability in AED, which takes a value between 0 and 1, is calculated using the decoder hidden state and the TCPGen output vector $\mathbf{h}^{\text{ptr}}_i$. In RNN-T, the TCPGen distribution is calculated for each combination of the encoder and the predictor step, and the query is computed from the corresponding encoder and predictor hidden states. Keys and values in RNN-T are computed from the predictor wordpiece embeddings. The generation probability for RNN-T is computed using the joint network output and the TCPGen output vector $\mathbf{h}^{\text{ptr}}_i$. To ensure that the probability of the null symbol in RNN-T is unchanged, $P^{\text{ptr}}_i(\varnothing|\mathbf{x}_{1:T}, y_{1:i-1})$ is set to 0 and the generation probability is scaled by $1-P^{\text{mdl}}_i(\varnothing|\mathbf{x}_{1:T}, y_{1:i-1})$ where $P^{\text{mdl}}_i$ is the original model distribution before interpolation. 

For more flexibility, an \textit{out-of-list} (OOL) token is included in $\mathcal{Y}^{\text{tree}}_i$ indicating that no suitable wordpiece can be found in valid wordpieces. To ensure that the sum of the final output distribution is 1, the generator probability is scaled as $\hat{P}^\text{gen}_i=P^\text{gen}_i(1-P^\text{ptr}(\text{OOL}))$. Therefore, the final output can be calculated as shown in Eqn. \eqref{eq:TCPGen_final}.
\vspace{-0.1cm}
\begin{equation}
    P(y_i) = P^{\text{mdl}}(y_i)(1-\hat{P}^\text{gen}_i) + P^{\text{ptr}}(y_i)P^\text{gen}_i,
    \label{eq:TCPGen_final}
    \vspace{-0.1cm}
\end{equation}
where conditions, $y_{1:i-1}, \mathbf{x}_{1:T}$, are omitted for clarity. $P^{\text{mdl}}(y_i)$ represents the output distribution from the standard end-to-end model, and $P^{\text{gen}}_i$ is the generation probability. 

\section{GNN encodings in TCPGen}
\label{sec:RecTCPGen}

This paper exploits a specific type of GNN, the tree-RNN structure for prefix-tree encoding in TCPGen. Specifically, at node $n_j$ which contains child nodes $n_1,...,n_k,...,n_K$, the vector representation of $n_j$ can be written as Eqn. \eqref{eq:treenet}.
\vspace{-0.1cm}
\begin{equation}
    \mathbf{h}^\text{tree}_{n_j} = f(\mathbf{W}_1\mathbf{y}_j + \sum_{k=1:K}\mathbf{W}_2\mathbf{h}^\text{tree}_{n_k}),
    \label{eq:treenet}
    \vspace{-0.1cm}
\end{equation}
where $\mathbf{h}^\text{tree}_{n_k}$ is the vector representation of node $n_k$, $f(\cdot)$ is an activation function which was ReLU in this paper, and $\mathbf{y}_j$ is the embedding vector of the wordpiece of node $n_j$. $\mathbf{W}_1$ and $\mathbf{W}_2$ are parameter matrices jointly optimised with the ASR system by allowing gradient back-propagation through $\mathbf{h}^\text{tree}_{n_k}$. In this way, each node recursively encodes information from its child nodes, such that the information of the entire branch rooted from it can be incorporated in the node encoding $\mathbf{h}^\text{tree}_{n_k}$. 
\begin{figure}[t]
    \centering
    \includegraphics[scale=0.3]{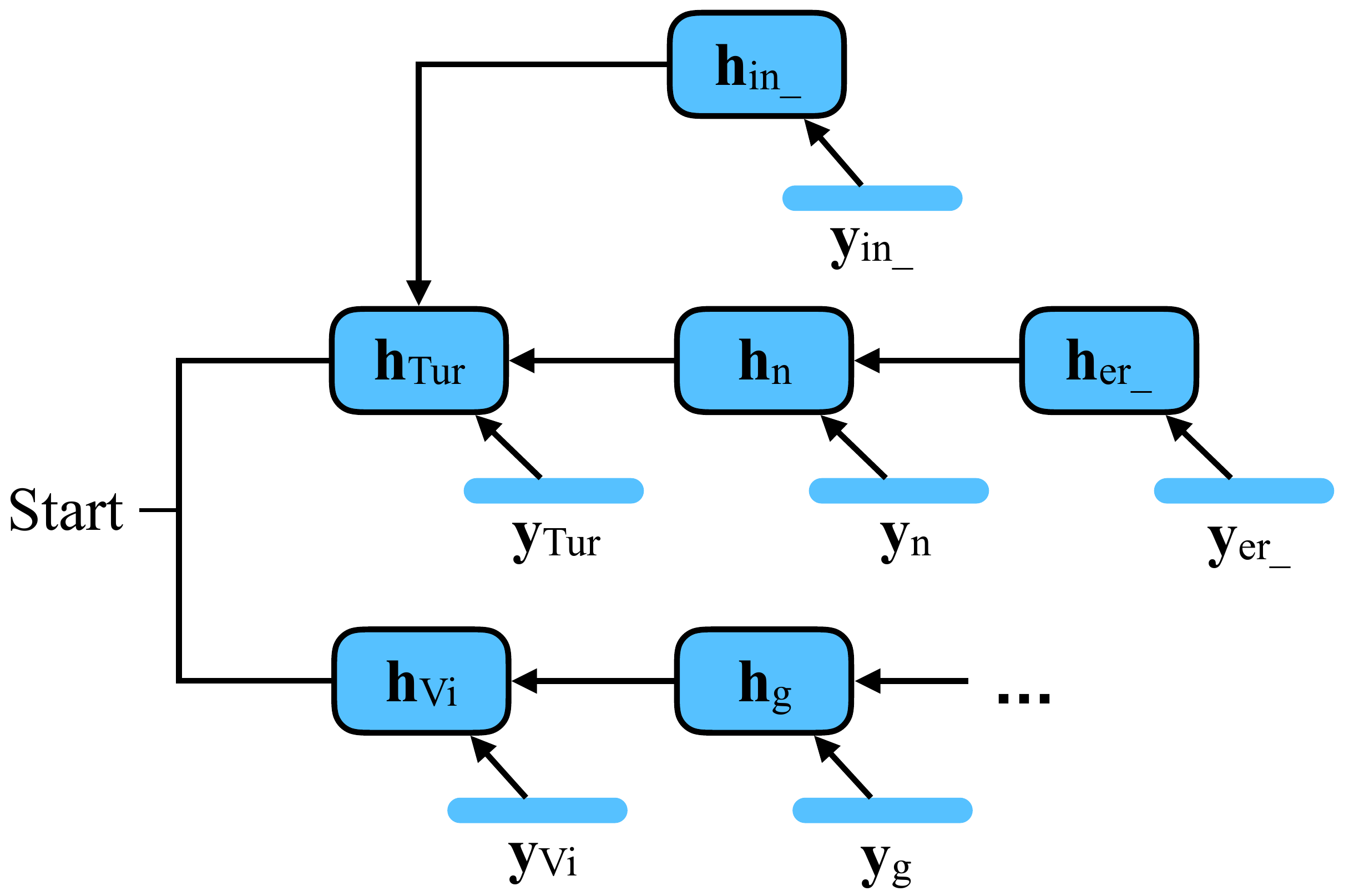}
    \vspace{-0.3cm}
    \caption{Illustration of tree-RNN computation over the prefix-tree representing the biasing list of 3 biasing words. A word end unit is denoted by \texttt{\_}.}
    \label{fig:recursivenet}
    \vspace{-0.3cm}
\end{figure}

\begin{figure*}[t]
\vspace{-0.3cm}
    \centering
    \includegraphics[scale=0.25]{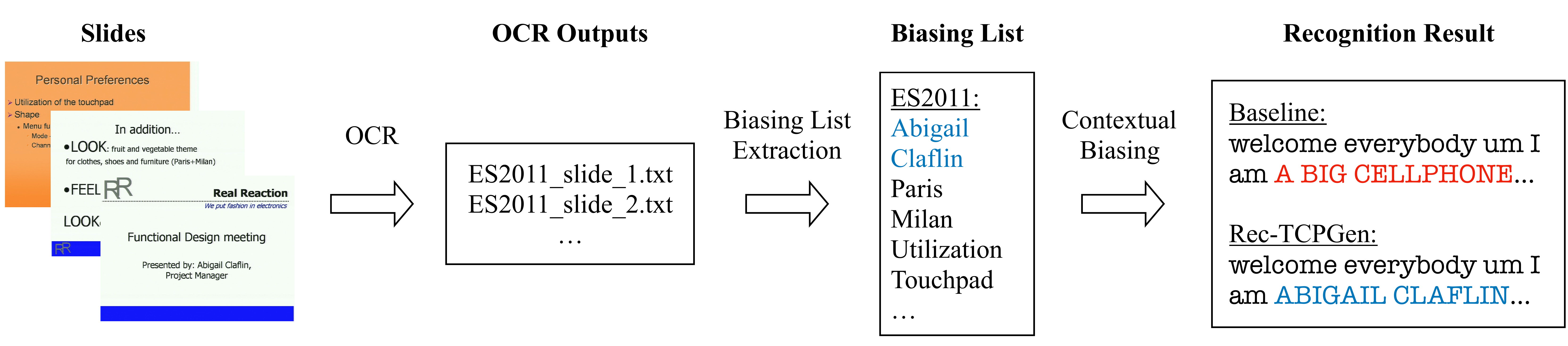}
    \vspace{-0.3cm}
    \caption{Illustration of the visual-grounded contextual ASR pipeline for meeting series ES2011 containing meetings ES2011[a-d].}
    \label{fig:amisetup}
    \vspace{-0.5cm}
\end{figure*}
The keys and values used to calculate the TCPGen distribution come from the encoding of nodes in the set of valid wordpieces, in place of wordpiece embeddings as shown in Eqn. \eqref{eq:TCPGeninlas_key}.
\vspace{-0.1cm}
\begin{equation}
    \mathbf{k}_j = \mathbf{W}^{\text{K}}\mathbf{h}^\text{tree}_{n_j} \hspace{1cm} \mathbf{v}_j = \mathbf{W}^{\text{V}}\mathbf{h}^\text{tree}_{n_j},
    \label{eq:TCPGeninlas_key}
    \vspace{-0.1cm}
\end{equation}
where $\mathbf{W}^{\text{V}}$ and $\mathbf{W}^{\text{K}}$ are parameter matrices. Fig. \ref{fig:recursivenet} shows an example of tree-RNN computation over the prefix-tree. Before the forward pass of the ASR model, the encoding of each node is calculated from the leaf nodes to the root recursively following Eqn. \eqref{eq:treenet}. Then, if, for example, the preceding output wordpiece is \texttt{Tur}, \texttt{in\_} and \texttt{n} form the set of valid wordpieces $\mathcal{Y}^\text{tree}_i$, and $\mathbf{h}^\text{tree}_\texttt{in\_}$ and $\mathbf{h}^\text{tree}_\texttt{n}$ are used to calculate the TCPGen distribution and $\mathbf{h}^\text{ptr}_i$. If \texttt{Turner} appears in the utterance to be recognised, the information of the entire word can be seen by TCPGen as early as in the encoding of \texttt{Tur}. Such lookahead functionality in the prefix-tree search allows a more accurate generation probability to determine when contextual biasing is needed.

The high efficiency of TCPGen in handling large biasing list is also retained when the tree-RNN is used. In training, with a large biasing list of 1000 words, TCPGen with a tree-RNN was three times slower than the standard AED or RNN-T model, with a negligible increase in space complexity. However, since the GNN encodings can be generated offline before the start of decoding once the biasing list is available, making the time and space complexity during inference close to the standard AED or RNN-T for biasing lists of thousands of words.

\section{Experimental setup}
\label{sec:setup}
\vspace{-0.1cm}
\subsection{Data}
Experiments were performed using the Lirbispeech audiobook data and the AMI meeting corpus with individual headset microphone (IHM) recordings. The Librispeech corpus \cite{librispeech} contains 960 hours of read English from audiobooks. The dev sets were held out for validation, and test-clean and test-other used for evaluation. The AMI meeting corpus \cite{ami} contains 100 hours of meeting recordings with 4-5 people, and was split into train, dev and eval sets. To show the effectiveness of applying contextual biasing to data in another domain with limited training resources, 10\% of the utterances from the AMI train set corresponding to 8 hours of audio were used to finetune the models trained on the Librispeech 960-hour data. There were 14 meetings in the dev set and 8 meetings in the eval set 
with slides. 
They were put together to form the slides test set for AMI experiments. Model input used 80-dimensional (-d) log-Mel filter bank features at a 10~ms frame rate concatenated with 3-d pitch features. SpecAugment \cite{specaug} with the setting $(W,F,m_F,T,p,m_T)=(40,27,2,40,1.0,2)$ was used without any other data augmentation or speaker adaptation.
\vspace{-0.1cm}
\subsection{Biasing list extraction}
Biasing lists for Librispeech experiments were arranged following \cite{DBRNNT}: the full rare word list containing 200k distinct words was obtained by removing the most common 5k words in the Librispeech LM training corpus. Rare words were defined as words belonging to this list. Biasing lists were then organised by finding words that belong to the full rare word list from the reference of each utterance and adding 1000 distractors. 

The visual-grounded contextual ASR pipeline for AMI using OCR output for slides is shown in Fig. \ref{fig:amisetup}. First, slides for each meeting series (e.g. ES2011 for ES2011[a-d]) were collected and passed through the Tesseract 4 OCR engine with long-short term memory (LSTM) models\footnote{https://github.com/tesseract-ocr/tesseract}. Distinct word tokens were then extracted from the OCR output text files, and those in the full rare word list that also appeared fewer than 100 times in the AMI train set were included in the biasing list for that specific meeting series. 
These meeting-specific biasing lists were then used for the recognition of all utterances in that meeting series. The size of the biasing lists ranged from 175 to 576 and the total number of word tokens covered by these biasing lists was 1,751 out of 112,110 word tokens (1.5\%), and hence had a small impact on the overall WER. However, as shown in the example in Fig. \ref{fig:amisetup}, these words were mostly highly valuable content words, and correct recognition was critical to understanding the utterance. 
Details of the meetings with slides and the extraction pipeline can be found in \cite{code}.

\vspace{-0.1cm}
\subsection{Model and evaluation metrics}
Systems were built using the ESPnet toolkit \cite{espnet}. Conformer-based \cite{conformer} AED and RNN-T were used in the experiments. The encoder structure which comprised of 16 conformer blocks with 512-d hidden state and 4-head attention was the same for both AED and RNN-T. AED used a 1024-d single-head location-sensitive attention and a 1024-d single-layer unidirectional LSTM decoder. RNN-T used a 640-d single-layer uni-directional LSTM predictor and a 640-d fully-connected layer joint network. The encoding size of each node in the prefix tree was 1024-d. SF with a target domain LM was also performed. For AMI experiments, the density-ratio LM SF approach following \cite{density_ratio} was performed. The LM for the target domain was a 2-layer 2048-d unidirectional LSTM-LM trained on the joint AMI and Fisher text corpus. The source domain LM was a single-layer 1024-d unidirectional LSTM-LM trained on the same 10\% AMI train set used for ASR model training. 

Systems were evaluated using WER and rare word error rate (R-WER) following \cite{DBRNNT, TCPGen}. R-WER is the total number of \textit{error} word tokens that belong to the biasing list divided by the total number of word tokens in the test set that belong to the biasing list. In contrast to \cite{MEM}, insertion errors were counted in R-WER if the inserted word belonged to the biasing list. The slides rare word error rate (R$_\text{s}$-WER) reported for the AMI experiments was calculated the same way as R-WER, but for the rare words in slides. Insertions of slides biasing words were included in R$_\text{s}$-WER. Furthermore, a speaker-by-speaker sign test for TCPGen with or without GNN encodings was performed to show the significance of improvements.

\section{Results}
\label{sec:result}

\subsection{Librispeech experiments}
\label{sec:libriexp}
\vspace{-0.1cm}

First, experiments were performed on the Librispeech data using the AED as shown in Table \ref{tab:aedlibri}. Compared to the baseline, WER, and in particular, R-WER reductions were achieved using TCPGen. Compared to the original TCPGen, using GNN encodings achieved a further R-WER reduction, enlarging the relative R-WER improvements from 37\% to 50\% on the test-clean set and from 32\% to 45\% on the test-other set. 

\begin{table}[h]
    \centering
    \begin{tabular}{lcccc}
    \toprule
    & \multicolumn{2}{c}{Test-clean(\%)} & \multicolumn{2}{c}{Test-other(\%)} \\
    System & WER & R-WER & WER & R-WER \\ 
    \midrule
    Baseline & 3.8 & 13.4 & 9.6 & 30.5 \\
    TCPGen & 3.3 & 8.4 & 8.3 & 20.7 \\
    ~~+ GNN enc. & \textbf{3.1} & \textbf{6.7} & \textbf{7.9} & \textbf{17.8} \\
    \bottomrule
    \end{tabular}
    \caption{WER and R-WER using AED on Librispeech test-clean and test-other sets. Baseline refers to the standard AED model.}
    \label{tab:aedlibri}
    \vspace{-0.3cm}
\end{table}

\begin{figure}[t]
    \centering
    \includegraphics[scale=0.3]{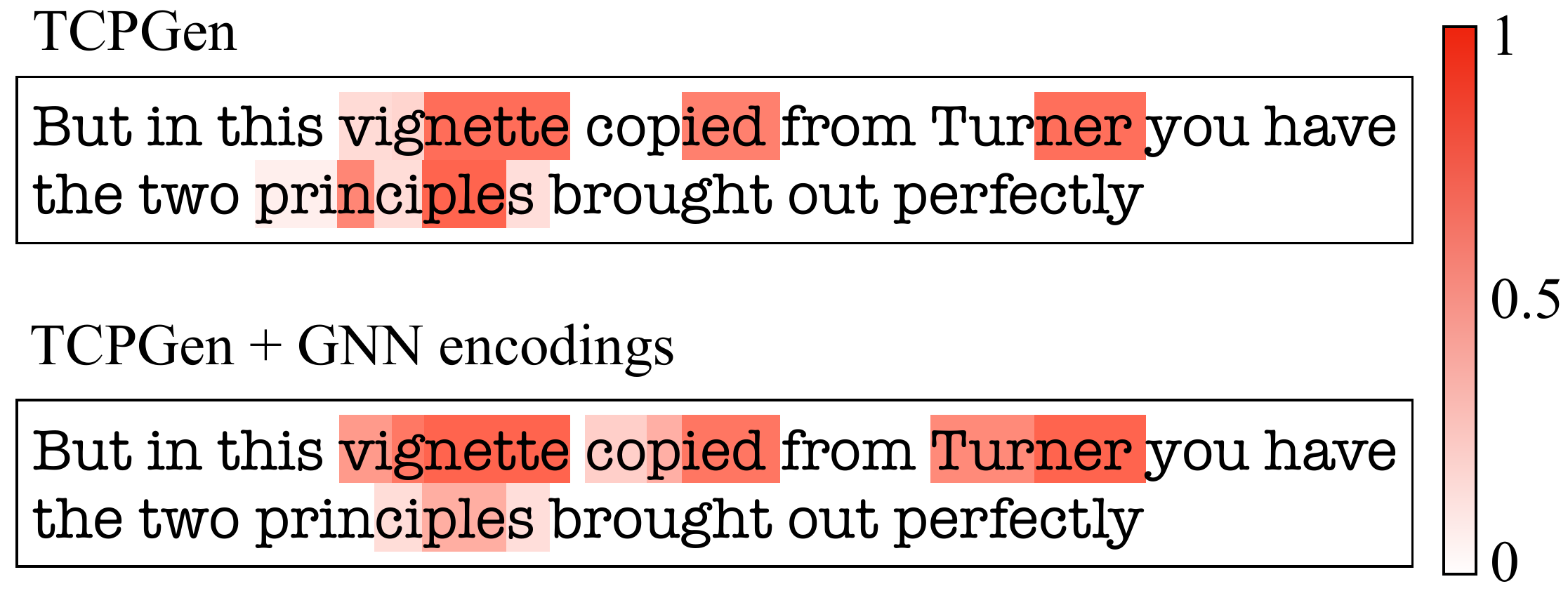}
    \vspace{-0.3cm}
    \caption{Heat map of the generation probability for each wordpiece in a recognised utterance illustrating the effect of lookahead using GNN encodings. Biasing words are \texttt{Vignette}, \texttt{copied} and \texttt{Turner}.}
    \vspace{-0.3cm}
    \label{fig:heatmap}
\end{figure}

To better illustrate how the lookahead functionality 
benefits the prediction of the generation probability in TCPGen, a heat map of $P^\text{gen}_i$ for each wordpiece in an example utterance is plotted in Fig. \ref{fig:heatmap}. The generation probabilities for the first several wordpieces of the biasing word were much higher with GNN encodings. This corroborates the claim in Sec.~\ref{sec:RecTCPGen} that since the tree-RNN allows information about wordpieces on the branches to be accessed at much earlier nodes, better generation probabilities were predicted than with the original TCPGen. 

\begin{table}[h]
    \centering
    \begin{tabular}{lcccc}
    \toprule
    & \multicolumn{2}{c}{Test-clean (\%)} & \multicolumn{2}{c}{Test-other (\%)} \\
    System & WER & R-WER & WER & R-WER \\ 
    \midrule
     Baseline  & 4.0 & 14.1 & 10.1 & 33.1 \\
     TCPGen & 3.4 & 8.9 & 8.8 & 22.2 \\
     ~~+ GNN enc. & \textbf{3.1} & \textbf{7.6} & \textbf{8.2} & \textbf{18.8} \\
    \bottomrule
    \end{tabular}
    \caption{WER and R-WER on Librispeech test-clean and test-other sets using RNN-T. Baseline refers to the standard RNN-T.}
    \label{tab:rnntlibri}
    \vspace{-0.3cm}
\end{table}
The effectiveness of GNN encodings was then verified for RNN-T on Librispeech data as shown in Table \ref{tab:rnntlibri}. Using GNN encodings further improved the relative R-WER reduction compared to the baseline standard RNN-T from 37\% to 46\% on the test-clean set and from 33\% to 43\% on the test-other set.

TCPGen also achieved \textit{zero-shot learning} of words that did not occur in training (referred to as OOV words) if incorporated in the biasing list, which was further improved with GNN encodings. There were 439 OOV word tokens in the combined test-clean and test-other set covered by biasing lists. For AED, the OOV WER, measured in the same way as R-WER but for OOV words, decreased from \textbf{73\%} to \textbf{40\%} using TCPGen, and then to \textbf{33\%} using GNN encodings. A similar reduction in OOV WER was found with RNN-T.

\subsection{AMI experiments}
\vspace{-0.1cm}
Experiments were performed on AMI for AED and RNN-T following the visual-grounded contextual biasing pipeline in Fig. \ref{fig:amisetup}, with results in Table \ref{tab:amiexp}. The baseline and TCPGen systems were all initialised from the corresponding models trained on the Librispeech 960-hour data from Sec. \ref{sec:libriexp}, and finetuned on 10\% of the AMI train set. Additionally, the AMI baseline was included, which used the same AED or RNN-T structure with the same set of wordpieces and trained on the full AMI train set. Compared to the AMI baseline, both finetuned standard models achieved a better WER with only 10\% of the full AMI train set.

\begin{table}[h]
    \centering
    \begin{tabular}{lcccc}
    \toprule
    & \multicolumn{2}{c}{AED (\%)} & \multicolumn{2}{c}{RNN-T (\%)} \\
    System & WER & R$_\text{s}$-WER & WER & R$_\text{s}$-WER \\
    \midrule
    AMI baseline & 23.6 & 56.3 & 26.5 & 58.0 \\
    \midrule
     Baseline &  22.2  & 51.2 & 25.7 & 52.6 \\
     TCPGen & 22.0 & 40.5 & 25.5 & 44.7 \\
     ~~+ GNN enc.  & \textbf{21.9} & \textbf{36.7} & \textbf{25.4} & \textbf{40.7}\\
     \midrule
     Baseline + SF & 21.1 & 45.5 & 24.3 & 46.5 \\
     TCPGen + SF & 20.9 & 34.2 & 24.1 &  37.7 \\
     ~~+ GNN enc. + SF  & \textbf{20.8} & \textbf{31.0} & \textbf{23.8} & \textbf{33.9}  \\
    \bottomrule
    \end{tabular}
    \caption{\%WER and \%R$_\text{s}$-WER on the AMI slides test set using AED and RNN-T finetuned on the 10\% of AMI train set. AMI baseline refers to the same AED or RNN-T system trained on full AMI train set only. SF denotes using shallow fusion.}
    \label{tab:amiexp}
    \vspace{-0.3cm}
\end{table}

The best WER and R$_\text{s}$-WER results without SF were obtained by GNN encodings in TCPGen for both AED and RNN-T on the 22 meetings of the AMI slides test set. Compared to the baseline, TCPGen achieved a 21\% relative R-WER reduction for AED, and 15\% relative R$_\text{s}$-WER reduction for RNN-T. These improvements increased to 28\% and 23\% for AED and RNN-T respectively using GNN encodings. The speaker-level sign test showed that R$_\text{s}$-WER reductions using GNN encodings compared to the original TCPGen were significant at $p<.05$. WER reductions were not significant for TCPGen but were significant for TCPGen with GNN encodings compared to the baseline at $p<.05$.

As the biasing lists were relatively small compared to the 1000 entries used in Librispeech, and hence the TCPGen distribution was sharp, insertions of biasing words occupied a non-negligible portion of the R$_\text{s}$-WER. To mitigate this issue, the density-ratio SF approach was used which also mitigated the domain mismatch between the training and test data. The SF factor for the target LM was 0.3 and the discounting factor for the source LM was 0.2 for all systems in Table \ref{tab:amiexp}. As a result, 
the relative R$_\text{s}$-WER reduction increased from 28\% to 32\% for AED, and from 23\% to 27\% for RNN-T when GNN encodings were used. The speaker-level sign test showed that R$_\text{s}$-WER reductions from using GNN encodings compared to the original TCPGen were all significant at $p<.05$ with SF. Similar improvements were also found using SF for TCPGen with GNN encodings on Librispeech data.

\section{Conclusions}
\label{sec:conclusion}

This paper proposed GNN encodings using tree-RNN in TCPGen for contextual speech recognition. Tree-RNN enables lookahead in prefix-tree search by encoding information about wordpieces on all branches rooted from each node. Experiments were performed on both Librispeech and AMI using both Conformer AED and RNN-T, with an audio-visual contextual ASR pipeline for AMI proposed as a realistic multi-modal evaluation setup. Consistent WER reductions were obtained using TCPGen with GNN encodings, with a significant reduction in R-WER and R$_\text{s}$-WER for biasing words compared to baselines. As a future direction, other effective GNN structures will be explored in TCPGen, such as the graph convolutional network (GCNs) \cite{gcn} and graph attention networks (GATs) \cite{gat}.

\bibliographystyle{IEEEtran}
\newpage

\begin{thebibliography}{9}
\bibitem{shallow_context_1}
I.~Williams, A.~Kannan, P.~Aleksic, D.~Rybach \& T.~Sainath,
\newblock {``Contextual speech recognition in end-to-end
neural network systems using beam search"},
 \newblock  {\em Proc. Interspeech}, Hyderabad, 2018.

\bibitem{shallow_context_2}
Z.~Chen, M.~Jain, Y.~Wang, M.~L.~Seltzer \& C.~Fuegen
\newblock {``End-to-end contextual speech recognition using class language models and a token passing decoder"},
 \newblock  {\em Proc. ICASSP}, Brighton, 2019.
 
\bibitem{shallow_context_3}
D.~Zhao, T.~Sainath, D.~Rybach, P.~Rondon, D.~Bhatia, B.~Li \& R.~Pang,
\newblock {``Shallow-fusion end-to-end contextual biasing"},
 \newblock  {\em Proc. Interspeech}, Graz, 2019.
 
\bibitem{deep_context_1}
G.~Pundak, T.~Sainath, R.~Prabhavalkar, A.~Kannan \& D.~Zhao
\newblock {``Deep context: {E}nd-to-end contextual speech recognition"},
 \newblock  {\em Proc. ICASSP}, Calgary, 2018.
 
\bibitem{deep_context_2}
Z.~Chen, M.~Jain, Y.~Wang, M.~L.~Seltzer \& C.~Fuegen
\newblock {``Joint grapheme and phoneme embeddings for contextual end-to-end ASR"},
 \newblock  {\em Proc. Interspeech}, Graz, 2019.

\bibitem{deep_context_3}
M.~Jain, G.~Keren, J.~Mahadeokar, G.~Zweig, F.~Metze \&
Y.~Saraf,
\newblock {``Contextual RNN-T for open domain ASR"},
 \newblock  {\em Proc. Interspeech}, Shanghai, 2020.
 
 \bibitem{deep_context_4}
U.~Alon, G.~Pundak \& T.~Sainath,
\newblock {``Contextual speech recognition with difficult negative training examples"},
 \newblock  {\em Proc. ICASSP}, Brighton, 2019.
 
 \bibitem{deep_context_5}
Z.~Chen, M.~Jain, Y.~Wang, M.~Seltzer \& C.~Fuegen
\newblock {``End-to-end contextual speech recognition using class language models and a token passing decoder"},
 \newblock  {\em Proc. ICASSP}, Brighton, 2019.

\bibitem{deepshallow}
D.~Le, G.~Keren, J.~Chan, J.~Mahadeokar, C.~Fuegen \& M.~L.~Seltzer
\newblock {``Deep shallow fusion for RNN-T personalization"},
 \newblock  {\em Proc. SLT}, 2021.
 
 \bibitem{DBRNNT}
D.~Le, M.~Jain, G.~Keren, S.~Kim, Y.~Shi, J.~Mahadeokar, J.~Chan, Y.~Shangguan, C.~Fuegen, O.~Kalinli, Y.~Saraf \& M.~L.~Seltzer
\newblock {``Contextualized streaming end-to-end speech recognition with trie-based deep biasing and shallow fusion"},
 \newblock  arXiv: {\em 2104.02194}, 2021.
 
\bibitem{dialogue_contextual}
Y.~Weng, S.~S.~Miryala, C.~Khatri, R.~Wang, H.~Zheng, P.~Molino, M.~Namazifar, A.~Papangelis, H.~ Williams, F.~Bell \& G.~Tur,
\newblock {``Joint contextual modeling for ASR correction and language understanding"},
 \newblock  {\em Proc. ICASSP}, Barcelona, 2020.
 
\bibitem{dialogue_contextual_2}
D.~Liu, C.~Liu, F.~Zhang, G.~Synnaeve, Y.~Saraf \& G.~Zweig,
\newblock {``Contextual language model adaptation for conversational agents"},
 \newblock  {\em Proc. Interspeech}, Hyderabad, 2018.
 
\bibitem{audiovisual_context}
G.~L.~Chao, C.~C.~Hu, B.~Liu, J.~P.~Shen \& I.~Lane
\newblock {``Audio-visual TED corpus: enhancing the TED-LIUM corpus with facial information, contextual text and object recognition"},
 \newblock  {\em Proc. UbiComp}, 2021.

 \bibitem{lm_pointer}
D.~Liu, C.~Liu, F.~Zhang, G.~Synnaeve, Y.~Saraf \& G.~Zweig,
\newblock {``Contextualizing ASR lattice rescoring with hybrid pointer network language model"},
 \newblock  {\em Proc. Interspeech}, Shanghai, 2020.
  
\bibitem{unsupervised_context}
Y.~M.~Kang \& Y.~Zhou
\newblock {``Fast and robust unsupervised contextual biasing for speech recognition"},
 \newblock  arXiv: {\em 2005.01677}, 2020.
 
 \bibitem{word_mapping}
R.~Huang, O.~Abdel-hamid, X.~Li \& G.~Evermann
\newblock {``Class LM and word mapping for contextual biasing in end-to-end ASR"},
 \newblock  {\em Proc. Interspeech}, Shanghai, 2020.
 
 \bibitem{pointer_1}
A.~See, P.~J.~Liu \& C.~D.~Manning
\newblock {``Get to the point: summarization with pointer-generator networks"},
 \newblock  {\em Proc. ACL}, Vancouver, 2017.
 
 \bibitem{e2e_aed_1}
W.~Chan, N.~Jaitly, Q.~V.~Le \& O.~Vinyals,
\newblock {``Listen, attend and spell: {A} neural network for large vocabulary conversational speech recognition"},
 \newblock  {\em Proc. ICASSP}, Shanghai, 2016.
 
 \bibitem{e2e_rnnt_1}
A.~Graves, A.~Mohamed \& G.~Hinton,
\newblock {``Speech recognition with deep recurrent neural networks"},
 \newblock  {\em Proc. ICASSP}, Vancouver, 2013.
 
\bibitem{TCPGen}
G.~Sun, C.~Zhang \& P.~C.~Woodland
\newblock {``Tree-constrained pointer generator for end-to-end contextual speech recognition"},
 \newblock  {\em Proc. ASRU}, Cartagena, 2021.
 
 \bibitem{MEM}
C.~Huber, J.~Hussain, S.~Stüker \& A.~Waibel
\newblock {``Instant one-shot word-learning for context-specific neural sequence-to-sequence speech recognition"},
 \newblock  {\em Proc. ASRU}, Cartagena, 2021.
 
\bibitem{neuralfst}
A.~Bruguier, D.~Le, R.~Prabhavalkar, D.~Li, Z.~Liu, B.~Wang, E.~Chang, F.~Peng, O.~Kalinli \& M.~L.~Seltzer
\newblock {``Neural-FST class language model for end-to-end speech recognition"},
 \newblock  {\em Proc. ICASSP}, Singapore, 2022.
 
\bibitem{contextselection}
M.~Han, L.~Dong, Z.~Liang, M.~Cai, S.~Zhou, Z.~Ma \& B.~Xu
\newblock {``Improving end-to-end contextual speech recognition with fine-grained contextual knowledge selection"},
 \newblock  {\em Proc. ICASSP}, Singapore, 2022.
 
 \bibitem{RNN1}
C.~Goller \& A.~Kuchler,
\newblock {``Learning task-dependent distributed representations by backpropagation through structure"},
 \newblock  {\em Proc. ICNN}, Washington, 1996.
 
  \bibitem{RNN2}
P.~Frasconi, M.~Gori \& A.~Sperduti
\newblock {``A general framework for adaptive processing of data structures"},
 \newblock  {\em IEEE Transactions on Neural Networks}, vol. 9, no. 5, pp. 768-786, 1996.
 
 \bibitem{RNNsemantic1}
X.~Chen, X.~Qiu, C.~Zhu, S.~Wu \& X.~Huang
\newblock {``Sentence modeling with Gated recursive neural network"},
 \newblock  {\em Proc. EMNLP}, Lisbon, 2015.
 
 \bibitem{RNNsemantic2}
T.~Zhang, M.~Huang \& L.~Zhao,
\newblock {``Learning Structured representation for text classification via reinforcement learning"},
 \newblock  {\em Proc. EMNLP}, Lisbon, 2015.
 
 \bibitem{RNNsentiment1}
H.~Sadr, M.~M.~Pedram \& M.~Teshnehlab 
\newblock {``A robust sentiment analysis method based on sequential combination of convolutional and recursive neural networks"},
 \newblock  {\em Neural Processing Letters}, Vol. 50, pp. 2745--2761, 2019.
 
 \bibitem{RNNNER}
P.~H.~Li, R.~P.~Dong, Y.~S.~Wang, J.~C.~Chou \& W.~Y.~Ma
\newblock {``Leveraging linguistic structures for named entity recognition with bidirectional recursive neural networks"},
 \newblock  {\em Proc. EMNLP}, Copenhagen, 2017.
 
 \bibitem{RNNsubword}
M.T.~Luong, R.~Socher \& C.~D.~Manning
\newblock {``Better word representations with recursive neural networks for morphology"},
 \newblock  {\em Proc. CoNLL}, Sofia, 2013.
 
 \bibitem{RNNsubword2}
M.~Nguyen, G.~H.~Ngo \& N.~F.~Chen
\newblock {``Hierarchical character embeddings: learning phonological and semantic representations in languages of logographic origin using recursive neural networks"},
 \newblock  {\em IEEE/ACM Transactions on Audio, Speech, and Language Processing}, Vol. 28, pp. 461--473, 2020.
 
 \bibitem{RNNtranslation1}
S.~Liu, N.~Yang, M.~Li \& M.~Zhou
\newblock {``A recursive recurrent neural network for statistical machine translation"},
 \newblock  {\em Proc. ACL}, Baltimore, 2014.
 
\bibitem{RNNtranslation2}
Y.~Kawara, C.~Chu \& Y.~Arase
\newblock {``Recursive neural network based preordering for english-to-japanese machine translation"},
 \newblock  {\em Proc. ACL-SRW}, Baltimore, 2018.
 
 \bibitem{transformer}
A.~Vaswani, N.~Shazeer, N.~Parmar, J.~Uszkoreit, L.~Jones, A.~N.~Gomez, L.~Kaiser \& I.~Polosukhin
\newblock {``Attention is all you need"},
 \newblock  {\em Proc. NIPS}, Long Beach, 2017.
 
 \bibitem{librispeech}
V. Panayotov, G. Chen, D. Povey \& S. Khudanpur,
\newblock {``Librispeech: {A}n ASR corpus based on public domain audio books"},
 \newblock  {\em Proc. ICASSP}, South Brisbane, 2015.
 
 \bibitem{specaug}
D.~S.~Park, W.~Chan, Y.~Zhang, C.~C.~Chiu, B.~Zoph, E.~D.~Cubuk \& Q.~V.~Le,
\newblock {``SpecAugment: {A} simple data augmentation method for automatic speech recognition"},
 \newblock  {\em Proc. Interspeech}, Graz, 2019.
 
\bibitem{density_ratio}
S.~Toshniwal, A.~Kannan, C.~C.~Chiu, Y.~Wu, T.~Sainath \& K.~Livescu,
\newblock {``A density ratio approach to language model fusion in end-to-end automatic speech recognition"},
 \newblock  {\em Proc. ASRU}, Singapore, 2019.
 
\bibitem{ami}
J.~Carletta, S.~Ashby, S.~Bourban, M.~Flynn, M.~Guillemot,
T.~Hain, J.~Kadlec, V.~Karaiskos, W.~Kraaij, M.~Kronenthal,
G.~Lathoud, M.~Lincoln, A.~Lisowska, I.~McCowan, W.~Post,
D.~Reidsma, \& P.~Wellner
\newblock {``The AMI meeting corpus: A preannouncement"},
 \newblock  {\em Proc. MLMI}, Bethesda, 2006.
 
\bibitem{conformer}
A.~Gulati, J.~Qin, C.~C.~Chiu, N.~Parmar, Y.~Zhang, J.~Yu, W.~Han, S.~Wang, Z.~Zhang, Y.~Wu \& R.~Pang,
\newblock {``Conformer: convolution-augmented transformer for speech recognition"},
 \newblock  {\em Proc. Interspeech}, Shanghai, 2020.
 
\bibitem{espnet}
S.~Watanabe, T.~Hori, S.~Karita, T.~Hayashi, J.~Nishitoba, Y.~Unno, E.~Y.~Soplin, J.~Heymann, M.~Wiesner, N.~Chen, A.~Renduchintala \& T.~Ochiai
\newblock {``ESPnet: {E}nd-to-end speech processing toolkit"},
 \newblock {\em Proc. Interspeech}, Hyderabad, 2018.
 
\bibitem{gcn}
T.~N.~Kipf \& M.~Welling,
\newblock {``Semi-supervised classification with graph convolutional networks"},
 \newblock {\em Proc. ICLR}, Toulon, 2017.
 
\bibitem{gat}
P.~Veličković, G.~Cucurull, A.~Casanova, A.~Romero, P.~Liò \& Y.~Bengio,
\newblock {``Graph attention networks"},
 \newblock {\em Proc. ICLR}, Vancouver, 2018.
 
 \bibitem{code}
\newblock {https://github.com/the-anonymous-bs/AMIslides\_biasing.git}

\end{thebibliography}

\end{document}